\documentclass[pra,twocolumn,showpacs,preprintnumbers]{revtex4}

\usepackage[dvips]{color}
\usepackage{graphicx}
\usepackage{epsfig}
\usepackage{dcolumn}
\usepackage{amssymb}
\usepackage{amsmath}
\usepackage[english]{babel}
\usepackage{mathrsfs}
\usepackage{bm}
\usepackage{lscape}

\newcommand{\ket}[1]{\left|#1\right\rangle}
\newcommand{\bra}[1]{\left\langle#1\right|}

\newcommand{\roundbraket}[1]{ \left( {#1} \right) }
\newcommand{\curlybraket}[1]{ \left\{ {#1} \right\} }

\def\*#1{\mathbf{#1}}
\def\ra{\rightarrow}

\begin{document}

\title{Strongly coupling a cavity to inhomogeneous ensembles of emitters : potential for long lived solid-state quantum memories}

\author{I. Diniz$^{1}$}
\author{S. Portolan$^{1}$}
\author{R. Ferreira$^{2}$}
\author{J.M. G\' erard$^{3}$}
\author{P. Bertet$^{4}$}
\author{A. Auff\`eves$^{1}$}\email{alexia.auffeves@grenoble.cnrs.fr}

\affiliation{$^{1}$ CEA/CNRS/UJF Joint team  ``Nanophysics and semiconductors,''
Institut N\'eel-CNRS, \\ BP 166, 25, rue des Martyrs, 38042 Grenoble Cedex 9, France}

\affiliation{$^{2}$ Laboratoire Pierre Aigrain, ENS/CNRS, 24 Rue Lhomond, F-75005 -
Paris, France}

\affiliation{$^{3}$ CEA/CNRS/UJF Joint team ``Nanophysics and semiconductors,''
CEA/INAC/SP2M, \\ 17 rue des Martyrs, 38054 Grenoble, France}

\affiliation{$^{4}$ Quantronics group, SPEC (CNRS URA 2464), IRAMIS, DSM, CEA, 91191
Gif-sur-Yvette, France}


\date{\today}

\begin{abstract}
We investigate theoretically the coupling of a cavity mode to a continuous distribution
of emitters. We discuss the influence of the emitters inhomogeneous broadening on the
existence and on the coherence properties of the polaritonic peaks. We find that their
coherence depends crucially on the shape of the distribution and not only on its width.
Under certain conditions the coupling to the cavity protects the polaritonic states from
inhomogeneous broadening, resulting in a longer storage time for a quantum memory based
on emitters ensembles. When two different ensembles of emitters are coupled to the
resonator, they support a peculiar collective dark state, also very attractive for the
storage of quantum information.
\end{abstract}

\maketitle

\section{Introduction}

%
Understanding the coupling between a cavity and an ensemble of emitters was motivated in
the early eighties by seminal demonstrations of cavity quantum electrodynamics (QED)
effects \cite{Serge}. First performed with atoms, these experiments were further
developed in solid state systems, starting with few semiconductor quantum wells coupled
to planar cavities \cite{Weisbuch}. The interest for this topic has been renewed in the
framework of quantum information, with proposals to use collections of emitters as
quantum memories for individual excitations. Indeed, ensembles of microscopic degrees of
freedom benefit from the collective enhancement of the interaction strength
\cite{Serge}, while possibly keeping the relaxation properties of a single emitter
\cite{Imamoglu}. This lead to a series of recent proposals where cold atoms
\cite{Verdu}, polar molecules \cite{Rabl} or electronic spins \cite{Weseberg, Imamoglu}
coupled to a superconducting cavity have been suggested as long-storage quantum memories
and optical interfaces. This problem also bears some analogy to the situation where a nuclear spin
ensemble is coupled to a single electronic spin \cite{Lukin}.
Following these proposals, recent experiments have demonstrated the strong coupling of a
resonator to a collection of electronic spins in a crystal \cite{Ong,Schuster}. However,
inhomogeneous broadening is always present in the solid state and may eventually limit
the performance of such a quantum memory.

In this paper, we study theoretically a cavity coupled to a continuous distribution of
inhomogeneously broadened emitters, in the low excitation regime.
In the ideal case where all the emitters have the same frequency, strong light-matter coupling leads
to the formation of two polaritonic modes separated by the so-called vacuum Rabi
splitting \cite{Serge2}. In the situation we aim to describe, the emitters bare frequencies are spread over a
range that can be larger than the cavity linewidth.
Our goal is to clarify the effect of inhomogeneous broadening on the former simple
picture in the ideal case, building on an early work by Houdr\' e et al \cite{Houdré}.
In the presence of inhomogeneous broadening, we also find polaritonic peaks. Surprisingly,
their relaxation properties are not only affected by the width of the emitters
distribution but also by its shape. We derive explicit formulas for the
polaritonic linewidths, showing in particular that provided the spectral density of
emitters in the wings of the distribution decays faster than a Lorentzian, the spectral
width will be dominated by the emitters {homogeneous} linewidth. We call this effect
\emph{cavity protection}. We solve exactly the dynamics of the coupled system, showing that, in this regime, the
two polariton states are well decoupled from the other emitters states. As a consequence,
{cavity protection} reduces very significantly the relaxation of an excitation,
when stored in one of the polariton states, which opens a promising path towards
solid-state quantum memories \cite{Lukin2, Zoltan}. We finally propose another potential
application of cavity protection, by considering a cavity coupled to two inhomogeneously
broadened ensembles of emitters. Indeed, this system supports a collective dark state,
which is particularly attractive for the storage of quantum information.

The paper is organized as follows. In Sec.\ref{sec:Model} we present the model leading
to Heisenberg equations in the low-excitation regime and obtain an expression for the
complex transmission of the cavity. This expression is analyzed in detail in
Sec.\ref{sec:TranmissionProperties} where we explore criteria for strong-coupling regime,
taking into account inhomogeneous broadening. The transmission pattern allows to introduce
the notion of cavity protection, whose physical origin is analyzed from two different perspectives in
Sec.\ref{sec:exactEingenV} and \ref{sec:DynamicalCount}.
Finally in Sec.\ref{sec:QM}, we study the potential of cavity protection in the framework of quantum memories. In particular, we discuss the possibility of exploiting a collective dark state to store and retrieve
quantum information with high fidelity.

\section{Model} \label{sec:Model}

\begin{figure}[htb]
\begin{center}
\includegraphics[width=8cm]{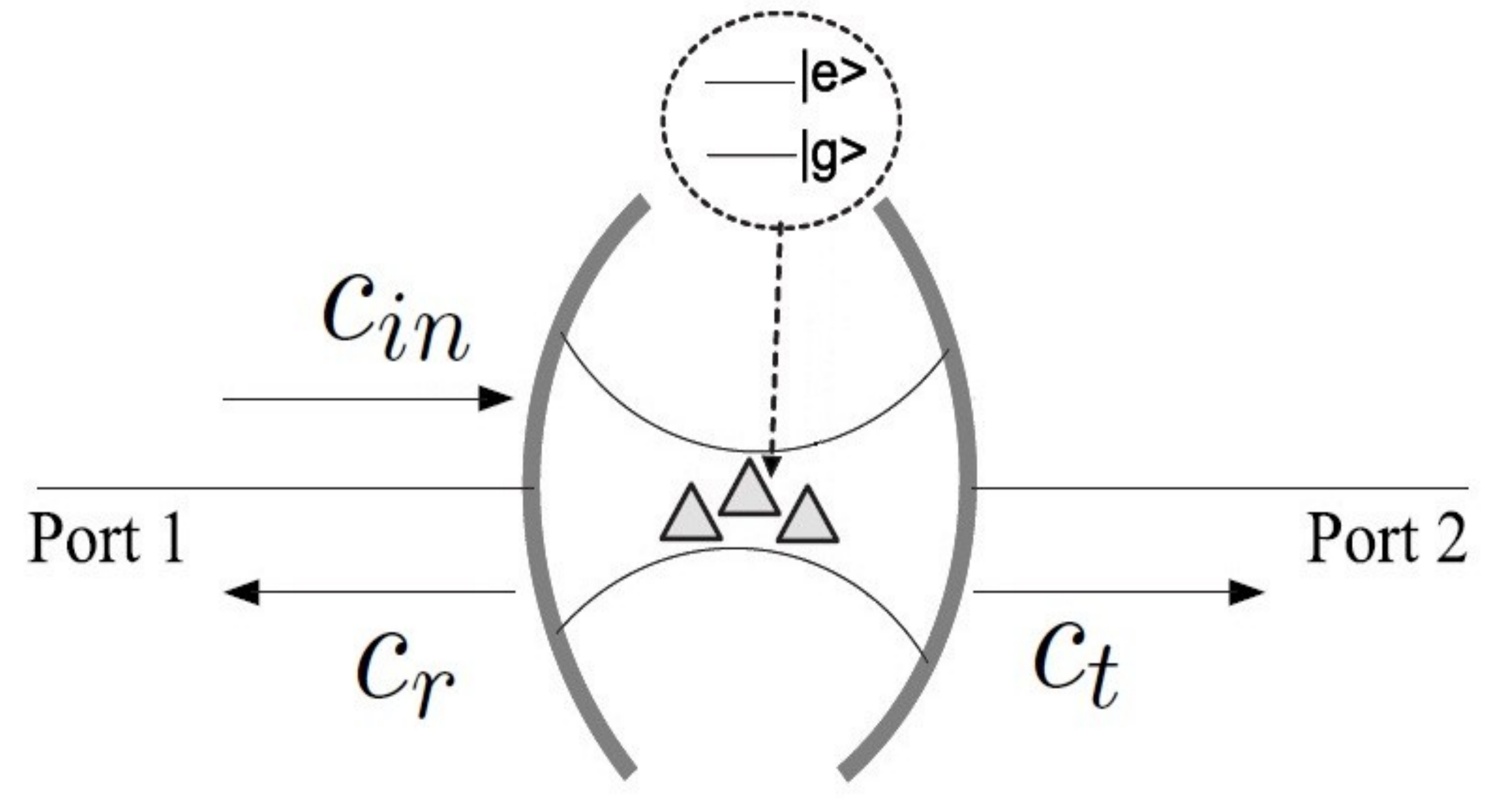}
\caption{Scheme of the emitters-cavity coupled system. The cavity frequency is
$\omega_0$. The cavity mode is coupled to the outside world via two ports labelled 1 and
2, the $k$-th two-level system has frequency $\omega_k$ and interacts with the cavity
mode with coupling constant $g_k$. } \label{figScheme}
\end{center}
\end{figure}

The system under study is pictured in fig.\ref{figScheme}. It consists in a cavity mode
$a$ of frequency $\omega_0$, that we shall define as the origin of frequencies, linearly
coupled with a strength $g_k$ to a distribution of $N$ two-level systems of frequencies
$\omega_k$ and damping rates $\gamma$. In the regime where the number of excitations is
small compared to the total number of emitters, each two-level emitter is properly modeled
by a bosonic mode $b_k$ (Holstein-Primakoff approximation). The total Hamiltonian writes
$H=H_{cav} + H_{em} + H_{int}$, with $H_{cav}=\hbar\omega_0 a^\dagger a$, $H_{em}=
\sum_k \hbar\omega_k b^\dagger_k b_k$ and $H_{int} = i \hbar\sum_k g_k (a^\dagger b_k -
b^\dagger_k a)$.

Using well-known input-output formalism \cite{Gardiner}, we define the external fields
$c_{in}$ (injected or pumping field), $c_r$ (reflected field) and $c_t$ (transmitted
field) that lead to the damping $\kappa$ of the intra-cavity field. We also consider
atomic losses $\gamma$, i.e. atomic emission in modes other than the cavity mode. The
Heisenberg equations are written in the frame rotating at the frequency $\omega$ of the probe,
yielding

\begin{equation}\begin{split}\label{input}
&\dot{a} = - \roundbraket{\kappa/2 + i(\omega_0 - \omega)}  a- \sqrt{\kappa/2} c_{in} + \sum_k g_k b_k  + f_a(t) \\
&\dot{b_k} = - \roundbraket{\gamma/2 + i(\omega_k - \omega)} s_k - g_k a + f_k(t) \\
& c_r = c_{in} + \sqrt{\kappa/2} a \\
& c_t = \sqrt{\kappa/2} a  \,
\end{split}\end{equation}

where $f_a(t)$, $f_k(t)$ are noise operators allowing the preservation of the
commutation relations.
From this set of equations, and as demonstrated in App.\ref{appendix:Dynamics}, it comes out that
the evolution of the system can be modelled with a generalized Hamiltonian $H_{eff}$
involving the respective complex emitters and cavity frequencies
$\tilde{\omega}_k=\omega_k-i\gamma/2$ and $\tilde{\omega}_0 = \omega_0 -i \kappa/2$.
Consequently, the system made of $N$ atoms coupled to a cavity appears to be equivalent to an ensemble of $N+1$ coupled leaky cavities,
and the problem reduces to the study of the classical evolution of the field in each cavity. This exact analogy is the basis of the model.
Taking the average value and solving analytically the set of equations in the steady state regime, we get the following expression
for the complex transmission of the cavity:

\begin{equation}\begin{split}
t(\omega) = \frac{\langle c_t \rangle}{\langle c_{in} \rangle} = %
\frac{-\kappa/2i}{  \tilde \omega_0 -  \omega -  \sum_k g^2_k / ( \tilde \omega_k -  \omega )} \, .%
\end{split}\end{equation}

We are interested in the very large number of emitters $N$, so we describe the emitters
as a continuous distribution with spectral density $\rho'(\omega')$ spread around its
central frequency $\omega_c$ and normalized to $1$. The full width at half maximum
(FWHM) is denoted $\Delta$, and is used to parametrize each distribution.  We replace
$g_k$ by $g(\omega', x)$ to account for the fact that the emitters at frequency
$\omega'$ can have a spread in the coupling constant, so that
\begin{equation}\begin{split}
  t(\omega) = \frac{-\kappa/2i}{\omega_0 - i \kappa/2 -  \omega - N \int d\omega' \frac{ \rho'(\omega') }{ \omega' - i\gamma/2 -  \omega } \int g^2(\omega',x) dx  } \, .%
\end{split}\end{equation}
The integration in $x$ can be carried out independently, yielding
\begin{equation}\begin{split}
   g_0^2(\omega') = \int g^2(\omega',x)  dx\, .%
\end{split}\end{equation}
Introducing the collective coupling constant $\Omega$ such as $\Omega^2 \rho(\omega)= N
\rho'(\omega) g^2_0(\omega)$ with $\rho(\omega)$ normalized to $1$,  we obtain

\begin{equation}\begin{split}  \label{eq:t(omega)in-out}
  t(\omega) = \frac{\kappa/2 i}{\omega -  \omega_0  + i \kappa/2  -  W(\omega) } \, , \\
\end{split}\end{equation}
with
\begin{equation}\begin{split}  \label{eq:Wdefinition}
  W(\omega) = \Omega^2\int_{-\infty}^\infty \frac{ \rho(\omega') d\omega'}{\omega - \omega' + i\gamma/2} .%
\end{split}\end{equation}

In the following we consider three different continua, namely a Gaussian, a
Lorentzian, and a rectangular distribution. Gaussian broadening is quite common in
nature, from Doppler-broadened lines in gases to e.g. size distributions in
ensembles of semiconductor nanocrystals \cite{Murray} and self-assembled quantum dots
\cite{Marzin}. Lorentzian distributions can be found in certain solid-state systems,
such as spin ensembles in dipolar interaction \cite{Hove} or dilute optically active
impurities in crystals \cite{Orth}. Finally, the rectangular distribution is a
prototypical example of finite bandwidth distribution. The results obtained in this case
can for instance qualitatively be applied to dilute ensembles of fluorescent molecules
in organic crystals \cite{Nicolet}. For these three distributions, we have obtained
analytical expressions for the function $W(\omega)$, which are detailed in Appendix
\ref{appendix:specificW}.

\section{Properties of the transmission function} \label{sec:TranmissionProperties}

In this section we discuss the properties of the transmission function
(eq.(\ref{eq:t(omega)in-out})), in the resonant case. First we recall some well-known results in the absence of inhomogeneous broadening
($\Delta = 0$). In that case, the distribution $\rho(\omega)$ is well described by a
Dirac delta function, leading to $W(\omega) = \Omega^2/(\omega+i\gamma/2)$, and the transmission
function has two poles $\lambda_\pm = \pm
\sqrt{\Omega^2-((\kappa-\gamma)/4)^2} + i{\displaystyle \frac{\kappa+\gamma}{4}}$ \cite{Claudio}.  Strong
coupling is reached if $\Omega\gg\kappa,\gamma$ and is manifested by the appearance of a
doublet in the transmission pattern located at $\pm \Omega$ (at first order in $\kappa/\Omega,\gamma/\Omega$). These two peaks are the spectral counterpart of the coherent and reversible exchange of
a quantum of energy between the cavity field and the symmetrical state $\ket{S}$ of the
emitters ensemble, defined as $\ket{S} = \Omega^{-1} \sum g_k b_k^\dagger \ket{0}$.
The transmission coefficient $t(\omega)$ is proportional to
the Fourier-Laplace transform of the field's amplitude in the cavity
initially fed with a single excitation  $\bra{1,G} e^{ -i H_{eff} t/\hbar}  \ket{1,G}$ (this result is demonstrated in Appendix \ref{appendix:Dynamics}, generalizing
ref.\cite{benjamin} and is also valid in the case where $\Delta>0$).  The
so-called collective Rabi oscillation takes place at the frequency $\Omega$ defined
above, which in that case simply equals $\Omega = g_0\sqrt{N}$, and is damped on a
timescale given by the finite linewidth of the peaks. In that temporal picture, strong
coupling is reached when the excitation is exchanged several times before
being lost in the environment.

\begin{figure}[hbt]
\begin{center}
\includegraphics[width=8cm]{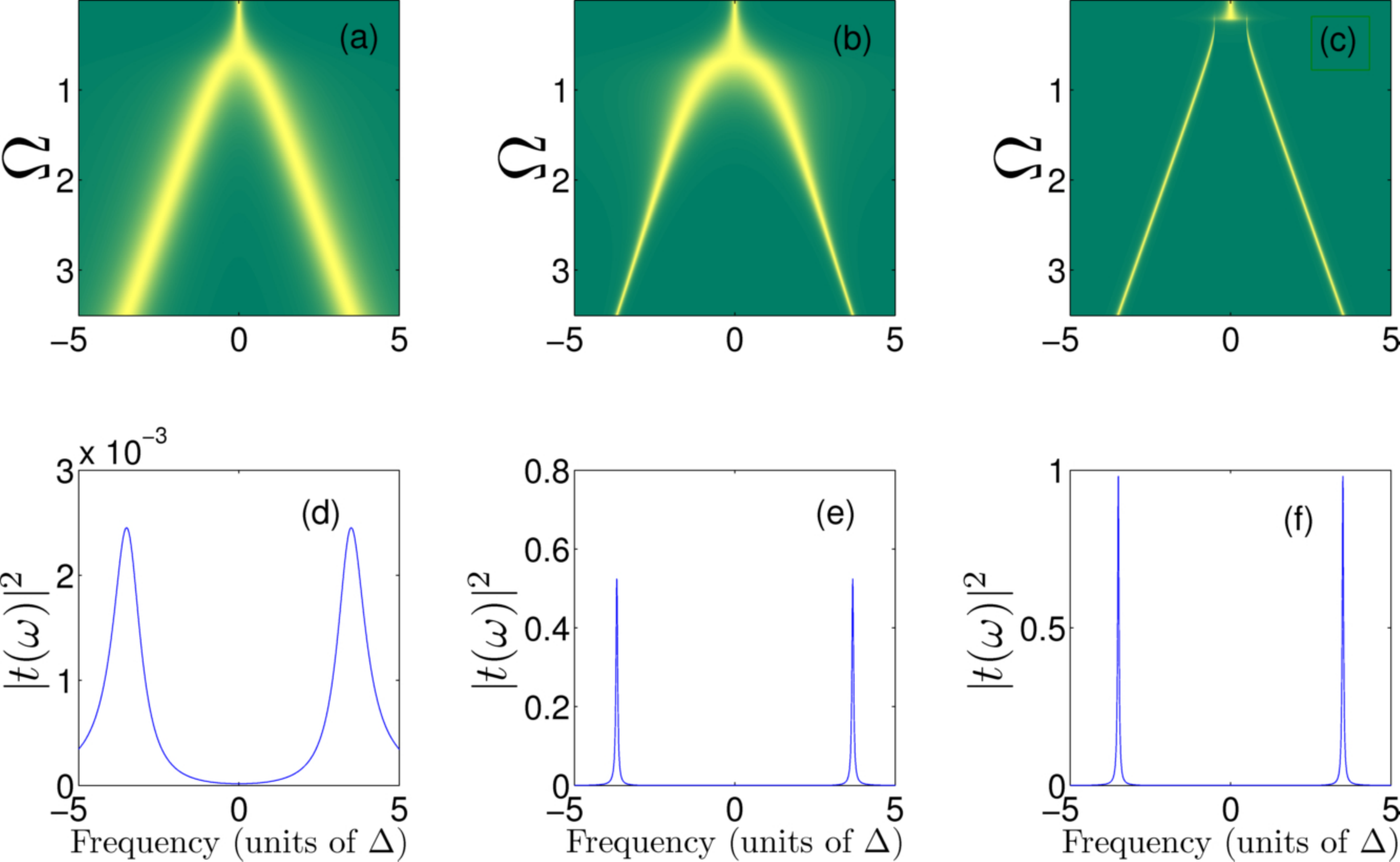}
\caption{(Color online). Transmission of a cavity resonantly coupled to a broad
distribution of emitters. (a,d) : Lorentzian; (b,e) : Gaussian; (c,f) : Rectangular. We
took $\Delta = 1$ MHz, $\kappa=0.1$ MHz, $\gamma=10^{-4}$ MHz. Bottom : $\Omega=3.5$
MHz. These values are typical of NV centers coupled to a superconducting resonator.} \label{fig:trans}
\end{center}
\end{figure}

We now study how the strong coupling features are modified by inhomogeneous broadening.
We have plotted the transmission in energy $|t(\omega)|^2$ for
$\Omega/\Delta$ ranging from $0$ to $3.5$ in fig.\ref{fig:trans}. To be only sensitive to the influence of
inhomogeneous broadening, we have kept $\kappa$ and $\gamma$ negligible with respect to
$\Omega$. We have considered the three types of distributions introduced in Sec.
\ref{sec:Model}, namely Lorentzian $(a)$, Gaussian $(b)$ and rectangular $(c)$. Whatever
the distribution, two peaks appear in the transmission pattern when $\Omega > \Delta$,
a signature of Rabi oscillation in the temporal domain. A first rough interpretation is that strong
coupling is reached when dephasing processes, that take place on a timescale
$\Delta^{-1}$, are slower than than energy exchanges, whose period still scales like
$\Omega^{-1}$. Note that the Rabi period is a collective quantity
involving all the emitters, even emitters which are not spectrally matched to the cavity
mode. This apparently puzzling feature had already been evidenced in \cite{Houdré} and
is due to the fact that the mode interacts with a collective state of the matter field.

Inhomogeneous broadening not only states a novel condition to fulfill to ensure strong
light-matter coupling. As it eventually accelerates the damping of Rabi oscillations, it
also leads to the broadening of the polaritonic peaks, as it clearly appears in
fig.\ref{fig:trans}. In particular, the shape of the emitters distribution has a
dramatic influence. An analytical expression for this width can be derived, in
perturbation with respect to the small parameter $\Delta/\Omega$ : namely, departing
from the strong coupling case in the absence of inhomogeneous broadening, we evaluate
how the poles of the transmission function are modified when $0<\Delta \ll \Omega$.  For
the sake of simplicity we consider the limit $\gamma=0$. The case of finite $\gamma$ is
studied in Appendix \ref{appendix:DevelopmentFinitegamma} in the limit $\gamma \ll
\Delta$, which corresponds to the experimental situations we aim to describe. Using the
Sokhatsky-Weierstrass formula in eq.(\ref{eq:Wdefinition}) we have
\begin{equation}\begin{split} \label{eq:Wnogamma}
\frac{W(\omega)}{\Omega^2} &= P\!\! \int_{-\infty}^\infty  \frac{\rho(\omega') d \omega' }{\omega - \omega' } %
-i \pi \rho(\omega)   \, .%
\end{split}\end{equation}

The modified poles of the transmission function are expected in the vicinity of $\pm \Omega$,
so that we develop the expression of $W(\omega)$ for $\omega \sim \Omega \gg \Delta$ :

\begin{equation} \begin{split} \label{eq:W(omega>Delta)}
W(\omega) &= \frac{\Omega^2 }{\omega}(1 + \mathcal{O}( \Delta^2 / \omega^2) )  - i \pi \Omega^2 \rho(\omega)  \, , %
\end{split}\end{equation}

yielding for the poles of the transmission function (at first order in $\kappa/\Omega$
and second order in $\Delta/\Omega$), $\lambda_\pm = \pm \Omega  + i\frac{\kappa+2\pi
\Omega^2 \rho(\Omega)}{4}$. Finally, keeping a finite $\gamma$ leads to the modified
expression for the full width at half maximum of the peaks :

\begin{equation} \label{eq:lambdapm}
\Gamma =( \kappa+\gamma+2\pi \rho(\Omega) \Omega^2)/2.
\end{equation}

Looking at eq.(\ref{eq:lambdapm}), it appears that in the strong coupling regime, the
polaritonic peaks remain located at $\pm\Omega$, but that inhomogeneous broadening adds
a contribution to their linewidth. This contribution writes $2 \pi
\Omega^2\rho(\Omega)$ and scales like the density of emitters at the real frequency of
the poles. This feature explains the sensitivity to the distribution shape that clearly appears in fig.\ref{fig:trans}.
The polaritonic linewidth decreases upon increasing $\Omega$, provided the distribution $\rho(\omega)$ decays faster than
$1/\omega^2$. The Lorentzian distribution is the limiting case for which the linewidth
tends towards a constant $\Delta$ : whatever the coupling, the polaritonic linewidth is
governed by inhomogeneous broadening. On the contrary, in the Gaussian and rectangular cases,
increasing the ratio $\Omega / \Delta$ allows to get rid of the influence of the
parameter $\Delta$, so that the width of the peaks only depends on the losses
of the cavity and of individual emitters. In the rectangular case, this ideal behavior is even reached
for finite values of the collective coupling strength $\Omega$ (while it remains a limit in the Gaussian case).
This effect, that we call {\it cavity protection},
leads to an enhanced lifetime of the Rabi oscillation and has interesting consequences for quantum information storage as we show in Section\ref{sec:QM}.

\section{Origin of peak broadening} \label{sec:exactEingenV}

Before focusing on applications opened by cavity protection, we give an interpretation of peaks broadening. This
amounts to understanding the damping of Rabi oscillations, which occurs even in the absence of any radiative
losses $\kappa=\gamma=0$. Our approach is based on a seminal paper of Fano \cite{Fano},
and consists in the diagonalization of the total Hamiltonian of the system $H = H_{cav} + H_{em} + H_{int}$.

In the absence of inhomogeneous broadening, preparing the system
in the initial state $\ket{1,G}$ gives rise to Rabi oscillations between the atoms and
the field. This state is a coherent superposition of two eigenstates of the
Hamiltonian, namely the polaritons $\ket{\psi_\pm^0}=\frac{1}{\sqrt 2} \ket{0,S} \pm i
\frac{1}{\sqrt 2}\ket{1,G}$, of energies $\pm \hbar \Omega$, where $\ket{S}$ is the symmetrical matter state defined in Section III.
Rabi oscillation is a quantum beat between these two components.
In particular, all other emitters states, which do not interact with the electromagnetic field and are usually called "dark states",
remain uncoupled.  The presence of inhomogeneous broadening strongly modifies the features of the emitters cavity coupling.
Introducing the continuous basis of bare emitters states $\ket{\omega}$ of energy $\hbar \omega$,
we write the matrix elements of $H$ as

\begin{equation} \begin{split} \label{eq:FourierFano}
&\bra{1,G}  H \ket{1,G} = \hbar \omega_0 \\ %
&\bra{ \omega'}  H \ket{1,G} = \hbar \Omega \sqrt{\rho(\omega')} \\ %
&\bra{ \omega'}  H \ket{ \omega} = \hbar \delta(\omega - \omega')  \omega  \, ,%
\end{split}\end{equation}
where the coupling is normalized per unit frequency.
An eigenvector $\ket{\psi_\omega}$ of $H$ with energy $\hbar \omega$ is searched under the form
\begin{equation} \begin{split}
\ket{\psi_\omega} = a(\omega) \ket{1,G} + \int d\omega' b(\omega,\omega') \ket{\omega'}
\, ,%
\end{split}\end{equation}

where the quantity $|a(\omega)|^2$ is normalized with respect to $\omega$. For distributions whose support is not bounded,
as it is the case for Lorentzian and Gaussian, the solution of the eigenvalue equation has been carried out by Fano in \cite{Fano}, yielding the normalized
eigenvectors:

\begin{equation} \begin{split} \label{eq:fanoeigenvec}
\ket{\psi_\omega} = \frac{ \sqrt{\rho(\omega)} \Omega \roundbraket{ \ket{1,G}  +
P\!\!\!\int d\omega' \frac{\sqrt{\rho(\omega')}  \Omega}{\omega - \omega'}
\ket{\omega'} } + C(\omega) \ket{\omega} }{\sqrt{C(\omega)^2 + (\pi \rho(\omega) \Omega^2 )^2 } } \, ,%
\end{split}\end{equation}
where $P\!\!\!\int$ stands for principal value and
\begin{equation} \begin{split}
C(\omega) = \omega  - \omega_0   -  \Omega^2 \; P\!\!\!\int d\omega' \frac{ \rho(\omega')}{\omega - \omega'}    \, .%
\end{split}\end{equation}

The amplitude of probability to find the excitation in the cavity mode can finally be written
\begin{equation} \begin{split} \label{eq:1G=F(a^2)}
 \bra{1,G} e^{ -i H t/\hbar}  \ket{1,G} &= \bra{1,G} e^{ -i H t / \hbar} \int d\omega' a^*(\omega') \ket{\psi_{\omega'}} \\%
    &= \int d\omega' |a(\omega')|^2  e^{-i\omega' t} \, .
\end{split}\end{equation}

It can easily be shown that $|a(\omega)|^2$ is proportional to the transmission coefficient in energy
$|t(\omega)|^2$ (namely, $|a(\omega)|^2=\Omega^2 \rho(\omega) \left|
\frac{t(\omega)}{\kappa/2}  \right|^2$ for $\gamma, \kappa \ra 0$), so that $|t(\omega)|^2$
corresponds to the Fourier transform of the occupation amplitude of the cavity mode.
As we have checked in Appendix \ref{appendix:DynamicsVsFano}, this result is completely consistent with the formalism of Laplace transform used in Section III
in the absence of external sources of losses.

This approach sheds new light on the transmission function studied in Section III, which directly reflects the overlap between the initial state $\ket{1,G}$ and
the continuum of eigenstates $\ket{\psi_\omega}$ of the Hamiltonian. The two peaks characteristics of the strong coupling regime show that this initial state is a coherent superposition of two wave packets,
reminiscent of the polaritons obtained when $\Delta=0$. As the eigenstates of the Hamiltonian form an infinite continuum, these wavepackets always have a finite width,
responsible for the damping of Rabi oscillations. Nevertheless, as it was shown above, increasing the collective coupling $\Omega$ may drastically change the shape of this overlap and
eventually, lead to the narrowing of the peaks for distributions $\rho(\omega)$ decaying faster than $\omega^{-2}$, a phenomenon that was defined above as cavity protection.

Distributions with a bounded support of width $\Delta$ (rectangular for example), provide an interesting limiting case where
cavity protection is almost perfect.
As a matter of fact, if $\Omega>\Delta$, the Hamiltonian eigenstates not only consist in a continuum $\psi_\omega$ lying within the
support of the distribution, but also in two discrete states $\ket{\psi_+}$ and $\ket{\psi_-}$, located around
$\omega=\pm \Omega$ (at first order in $\Delta/\Omega$), corresponding to the polaritons $\ket{\psi_+^0}$ and $\ket{\psi_-^0}$
when $\Delta = 0$. The initial state $\ket{1,G}$ mostly overlaps with these two eigenstates, making
the problem similar to the case of standard Rabi oscillations in the absence of inhomogeneous broadening.
In particular, if $\rho(\omega)$ is rectangular, the overlap of $\ket{1,G}$ with the discrete states equals ${\cal C}=1 - (1/8)
( \Delta / \Omega )^2 $, giving rise to Rabi oscillations of infinite duration characterised by a contrast ${\cal C}$.

To conclude this part, we emphasize that the total damping rate $\Gamma =( \kappa+\gamma+2\pi \rho(\Omega) \Omega^2)/2$
evidenced in Section III shows contributions of essentially different nature. The first type, related to $\kappa$ and $\gamma$, is due to the irreversible
loss of the excitation in the environment of the cavity or the emitters. The second type, related to $\pi \rho(\Omega) \Omega^2$, is Hamiltonian and thus reversible in principle
with CRIB experiments. It is due to the interaction of the cavity with a continuum of emitters, leading to progressive dephasing of Rabi oscillations.

\section{Open system approach} \label{sec:DynamicalCount}

\begin{figure}[ht]
\begin{center}
\includegraphics[width=8cm]{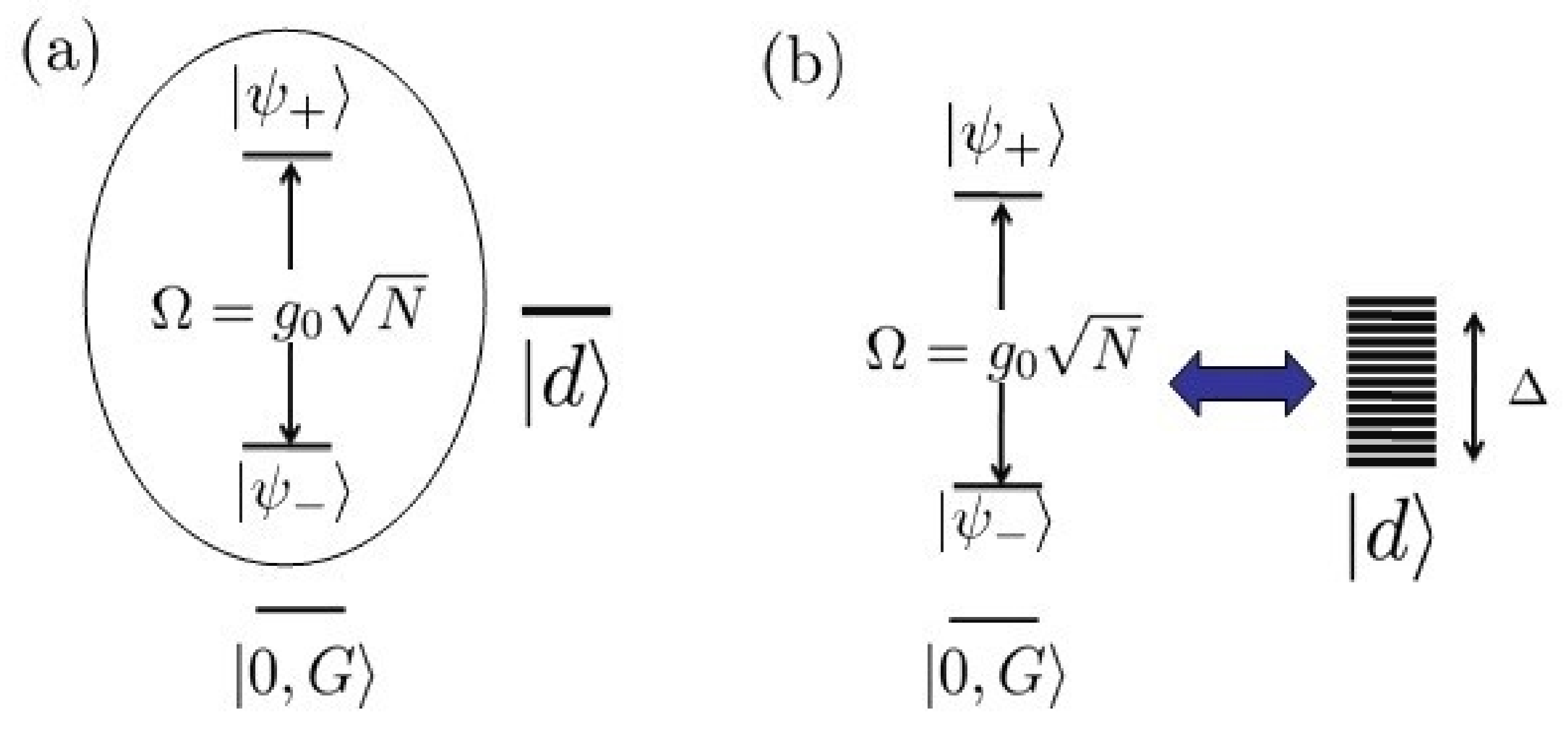}
\caption{(Color online). Schematic diagram of the open system approach to inhomogeneous
broadening. (a): $\Delta=0$, the states $\ket{\psi_\pm}$ are isolated from the
degenerate dark states $\ket{\omega}$. (b): $\Delta \neq 0$, the states $\ket{\psi_\pm}$
are coupled to the $\ket{\omega}$ states, which are non-degenerate in this case, with a
coupling strength proportional to $\Delta$ .} \label{fig:opensystem}
\end{center}
\end{figure}

The approach developed in Section IV gives an interpretation of the peaks
broadening within a Hamiltonian formalism. In this part, we adopt another point
of view based on quantum open systems.
As it was exposed above and pictured in fig.\ref{fig:opensystem}, in the absence of inhomogeneous broadening,
the symmetrical state $\ket{S}$ is decoupled from the dark states. The excitation initially injected in the cavity mode remains thus trapped
in the "small system" consisting in the two polaritons $\ket{\psi_+^0}$ and $\ket{\psi_-^0}$.
When inhomogeneous broadening is switched on, the symmetrical state couples to the dark states, which appear as an environment
in which the excitation can decay. Broadening of the polaritonic peaks can be attributed to the decoherence induced
by the bath of dark states. This picture is inforced by the computed expression for the width of the transmission peaks, $\Gamma=2\pi \Omega^2 \rho(\Omega)$,
which could be interpreted as a natural linewidth for polaritons "dressed" by the environment of dark states.
Nevertheless, the analogy should be used with caution, as the coupling with the bath is not
Markovian. This naive picture has still the advantage to give an intuitive insight on cavity protection, which is nothing
but energetically decoupling the polaritons from the bath of dark states, as initially suggested in \cite{Lukin2}.

\begin{figure}[ht]
\begin{center}
\includegraphics[width=9cm]{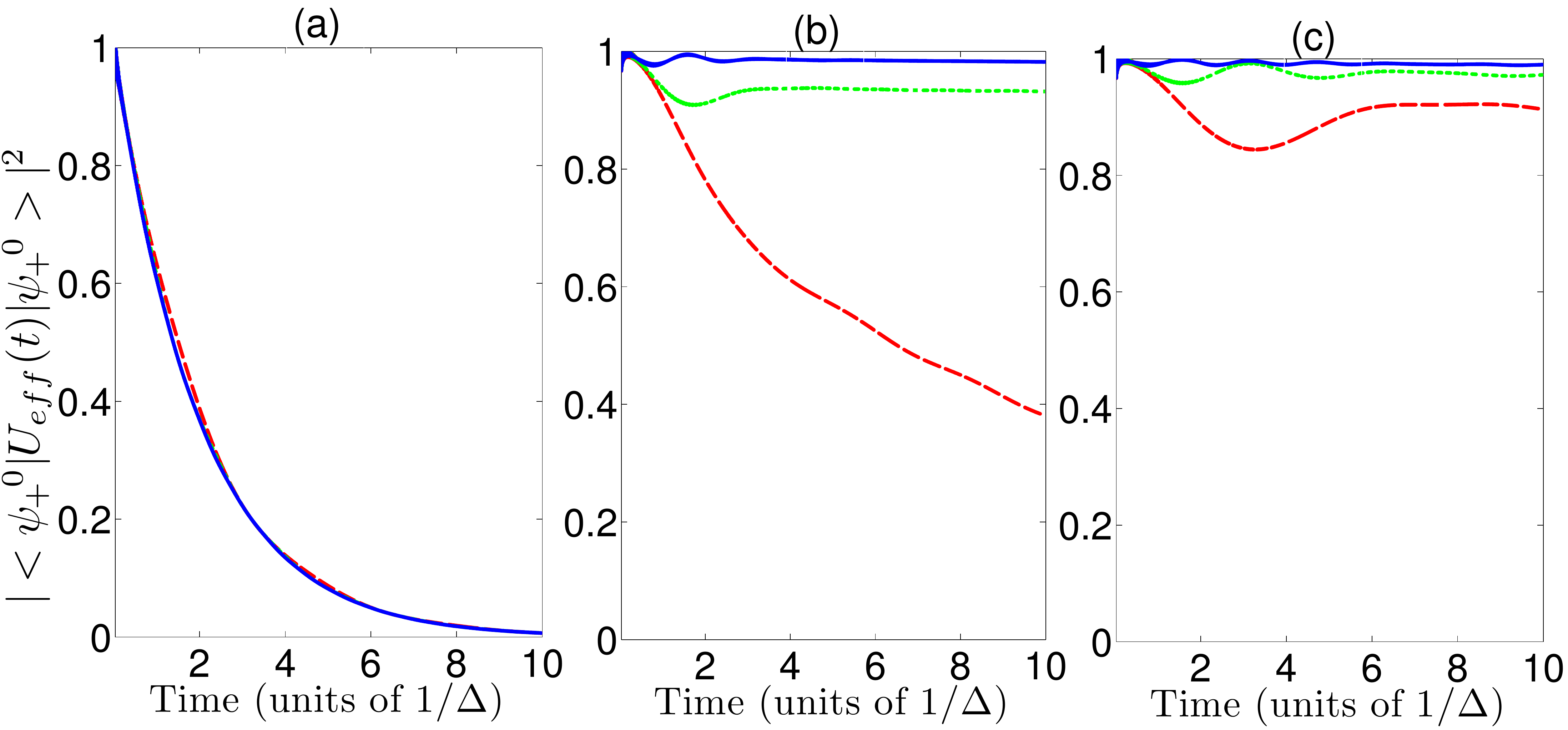}
\caption{(Color online). Probability to recover an excitation initially stored in the
state $\ket{\psi_+}$ after a time $t$. We took $\Delta=1$ MHz, $\kappa=\gamma=0$. (a) :
Lorentzian ; (b) : Gaussian ; (c) : Rectangular. Red Dashed line : $\Omega=1$ MHz; Green
Dotted line : $\Omega=2$ MHz; Blue line :  $\Omega=4$ MHz. } \label{decay}
\end{center}
\end{figure}

To study the dynamics of the polaritonic relaxation, we have exactly computed the evolution of the state of the system initially prepared in $\ket{\psi_+^0}$,
for different values of the collective coupling strength $\Omega$, and for the three types of distribution, keeping the same FWHM $\Delta=1$MHz.
We have plotted in fig.\ref{decay} the probability $|\bra{\psi_+^0} e^{-iH_{eff}t} \ket{\psi_+^0}|^2$ of finding the excitation in the polariton, as a function of time. For the sake of clarity,
we have neglected again the losses $\kappa=\gamma=0$ (realistic values are considered below).
As it can be seen in the figure, if the distribution is Lorentzian, the excitation exponentially decays in the environment, whatever the coupling $\Omega$, on a typical
timescale $\Delta^{-1}$. This is consistent with the spectral study performed in Section III, where the width of the polaritonic peaks does not depend on the coupling
with the cavity. On the contrary, the effect of cavity protection can be observed on the two other distributions. Damping is strongly inhibited as soon as $\Omega>\Delta$
if the distribution is Gaussian, but is always present whatever the coupling, which is the counterpart of the finite linewidth of the transmission peaks. Finally,
in the case of a rectangular distribution, two timescales are visible. The initial state $\ket{\psi^0_+}$ mostly overlaps with the discrete state $\ket{\psi_+}$
defined above, but also with the continuum of eigenstates $\ket{\psi_\omega}$. The coherent superposition of the continuum of frequencies is damped on a short timescale $\Delta^{-1}$,
so that the probability quickly converges towards the quantity $|\bra{\psi_+^0}\psi_+\rangle|^2$, which also scales like $(\Delta/\Omega)^2$.

\section{Application to quantum memories}  \label{sec:QM}

The previous Sections establish that for distributions allowing cavity protection, increasing the collective coupling $\Omega$
dramatically increases the potential storage time of one excitation in the polaritonic states,
as energetic decoupling from the dark states is more pronounced. In particular, this storage time becomes insensitive to dephasing processes induced by inhomogeneous broadening.
This allows to treat an inhomogeneous distribution as an effective oscillator of ground state $\ket{G}$ and first excited state $\ket{S}$, that
benefits from the collective coupling $\Omega$ to the cavity and whose relaxation
properties are solely governed by individual emitter properties $\gamma$.
As a consequence, cavity protection opens the path to the implementation of long lived solid-state quantum memories,
by exploiting ensembles of microscopic degrees of freedom, whose coherence times are remarkable.
In this Section we use our modelling to estimate the performances of two such types of quantum memories.

\subsection{Quantum memory based on dispersive coupling}

Here we evaluate the potential of a broad ensemble of emitters dressed by a cavity
mode for quantum information storage. The coupling should be dispersive to freeze Rabi
oscillations between the mode and the atoms. This system offers an interesting situation where
information has to be protected against two types of losses : the cavity losses, which are more
critical when the mode and the distribution of emitters are on resonance, and the losses
in the dark states, which on the contrary, are all the weaker as the atoms-cavity detuning is smaller.
The atoms-cavity detuning is thus the result of a tradeoff, and can be optimized with our modelling, as we show below.

\begin{figure}[ht]
\begin{center}
\includegraphics[width=8.8cm]{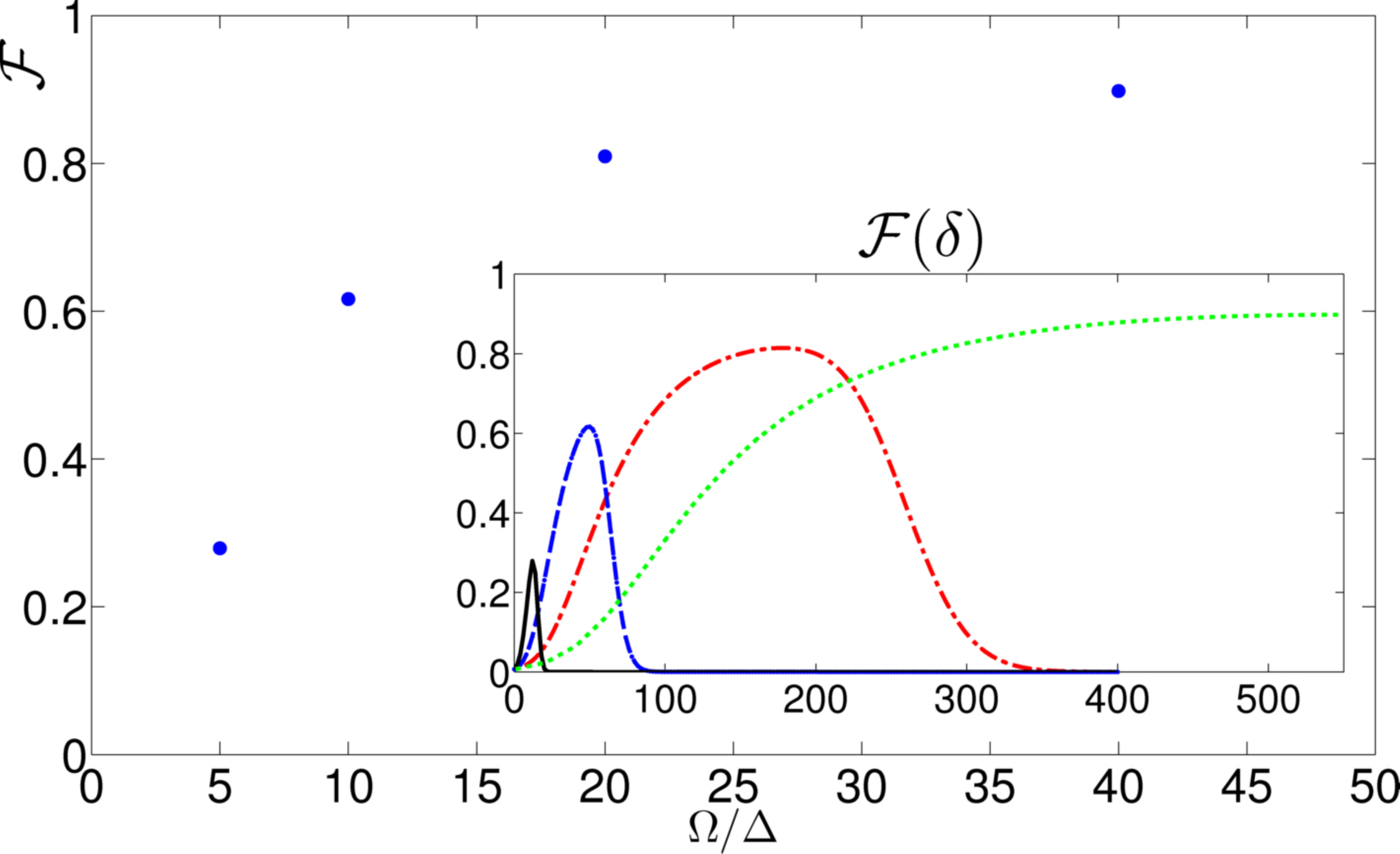}
\caption{(Color online). Maximized fidelity ${\cal F}$ of regaining the excitation
initially stored in the state $\ket{\psi_+}$ after $\tau=$ 10 cavity lifetimes, as a
function of $\Omega/\Delta$. We took $\Delta=1$ MHz, $\kappa=0.1$ MHz, $\gamma=10^{-4}$
MHz. Inset : same quantity ${\cal L}$ as a function of detuning $\delta$, after $\tau$.
Green dotted line : $\Omega=40$ MHz ; Black solid line : $\Omega=20$ MHz; Blue dashed
line : $\Omega=10$ MHz; Red dotted line : $\Omega = 5$ MHz.} \label{loss}
\end{center}
\end{figure}

The protocol of the quantum memory is the following. First, the detuning $\delta$ between the mode and the center of the distribution
is slowly swept from $-\infty$ to a finite positive value, thus adiabatically mapping the quantum state of
the cavity mode onto the emitter's ensemble : $(\alpha \ket{0}+\beta \ket{1})\ket{G}
\rightarrow \ket{0}(\alpha \ket{G} + \beta\ket{\psi^0_+(\delta)})$. We have introduced the
dressed state $\ket{\psi^0_+(\delta)}=\cos(\theta/2)\ket{0,S}+ i \sin(\theta/2)\ket{1,G}$, and the mixing angle $\cot(\theta)=\delta/ (2 \Omega)$. The transfer of the
excitation should be realized on a timescale longer than the Rabi period, but shorter
than $\Delta^{-1}$ so that no dephasing mechanism affects the process, this can be
achieved under strong coupling as in this case $\Omega \gg \Delta$ .
The expected fidelity ${\cal F}(t)$ of such a quantum memory can be exactly computed with the present
model; in particular, in the case where a single photon state is stored ($\beta=1$), we
get the simple expression ${\cal F}= |\bra{\psi^0_+(\delta)} e^{-iH_{eff}t}
\ket{\psi^0_+(\delta)}|^2$. We have plotted this quantity in fig.\ref{loss}. As explained above, ${\cal F}$
must be optimized by properly choosing the detuning $\delta$, which should be low enough to
maintain cavity protection, and high enough to reduce the sensitivity to cavity losses,
which typically scale like $\kappa (\Omega/\delta)^2$. The maximal detuning $\delta_M$ leading
to an efficient protective energy gap is $\Omega^2/\delta_{M}\sim \Delta$\cite{Zoltan}. This condition induces an optimal reduction
of the cavity losses by a factor of $(\Omega/\Delta)^2$.

The trade-off in the detuning clearly appears in the inset of fig.\ref{loss}, where we have plotted ${\cal F}$, as a function of the
detuning $\delta$, after ten cavity lifetimes, for different values of the ratio
$\Omega/\Delta$. We have used standard parameters for circuit QED technology \cite{Ong}. As it appears
in the figure, a quantum memory based on a Gaussian distribution of emitters of
linewidth $\Delta=1$ MHz, strongly coupled to a cavity of width $\kappa=0.1$ MHz with a
strength $\Omega =40$ MHz would yield a typical fidelity of $90\%$ after $100$ $\mu$s, a
remarkable storage time compared to the lifetime of the cavity mode ($10$  $\mu$s) and
the typical dephasing time of the ensemble ($1$ $\mu$s).

\subsection{Quantum memory based on two emitters distributions}

We focus now on a second type of quantum memory, based on two
distributions of emitters allowing cavity protection, respectively detuned by $+\delta$ and $-\delta$ with respect
to a cavity. Note that the case of a mode coupled to two such {\it discrete} emitters of ground and excited states $\ket{g_i}$ and $\ket{e_i}$ is exactly solvable, the poles of the transmission revealing
the complex eigenfrequencies of the system \cite{Fink}. In particular, when the emitters are on resonance with the mode ($\delta=0$),
the antisymmetrical state $(\ket{e_1,g_2}-\ket{g_1,e_2})/\sqrt{2}$ is not coupled to the electromagnetic field. This dark state is naturally protected against spontaneous emission in the cavity,
a property that can be used to store quantum information during a typical timescale given by the atomic dephasing time. Note that for artificial atoms like superconducting
qubits or quantum dots this time can be quite short, which is a severe drawback for quantum computation on chip.
Here we suggest an experiment to prepare and exploit this dark state as a quantum memory, in the case where the discrete emitters are replaced by broad assemblies of atoms.
This proposal allows to benefit from the collective atoms-cavity coupling, while the storage time now corresponds to the dephasing time of individual emitters, and is thus potentially
quite long. Note that this idea is typical of the so-called hybrid circuits technology \cite{Verdu, Rabl, Imamoglu, Weseberg}.

\begin{figure}[ht]
\begin{center}
\includegraphics[width=9cm]{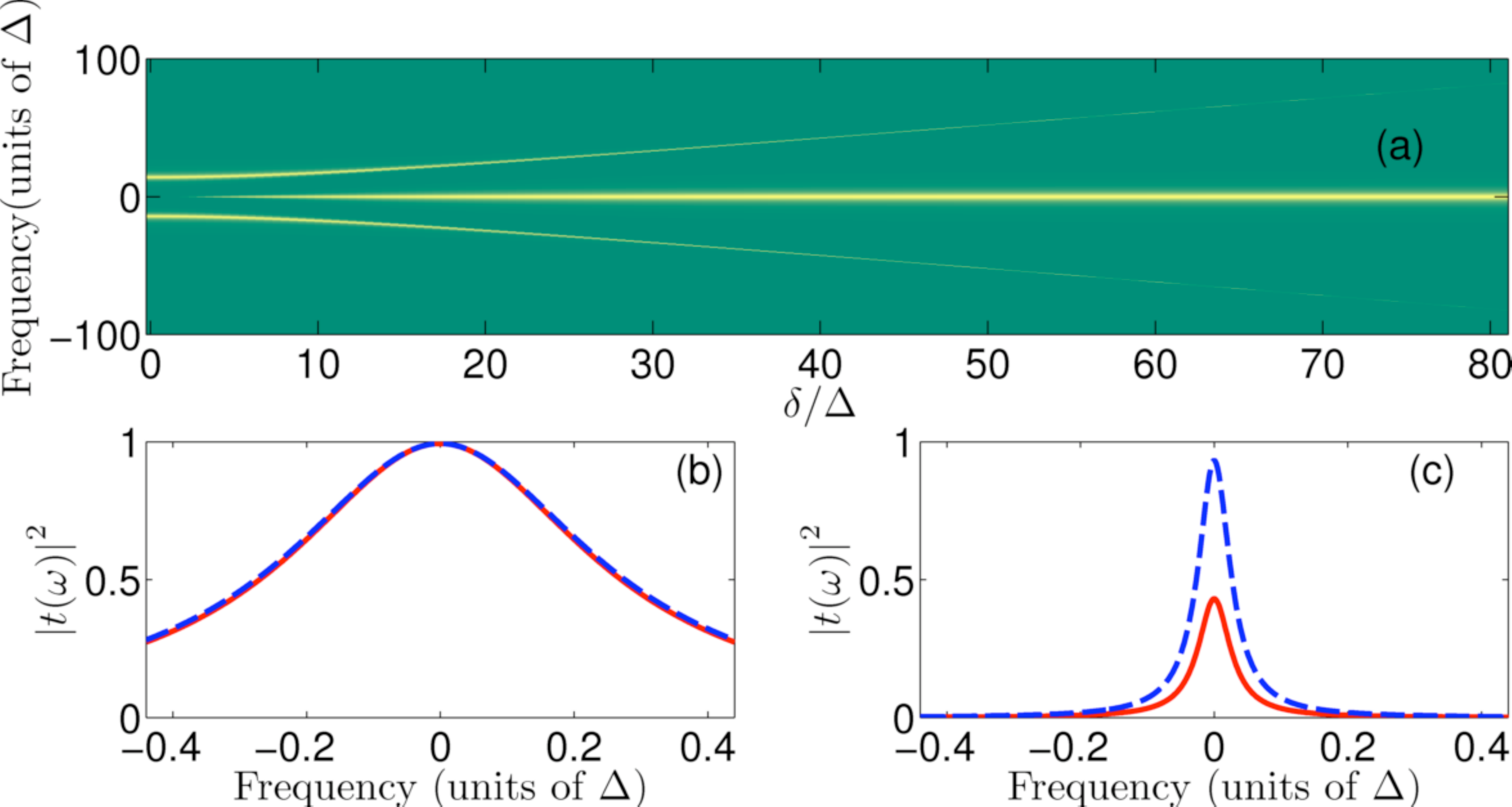}
\caption{(Color online). (a) : Transmission of a cavity coupled to two Gaussian
distributions of emitters, each detuned by $+\delta$ and $-\delta$ from the cavity
frequency, $\delta$ is swept from $0$ to $8$ MHz. We took $\Omega=1$MHz,
$\Delta=0.1$MHz, $\kappa=0.5$MHz, $\gamma=10^{-4}$MHz. (b) : Focus on the central peak
with $\delta=0.5$ MHz. Solid red line : Gaussian profile. Blue dashed line : two
emitters of homogeneous linewidth $\gamma$. (c) : $\delta=0.15$ MHz.} \label{dark}
\end{center}
\end{figure}

First we have checked the validity of the effective model if two ensembles are coupled
to the cavity. We have plotted in fig.\ref{dark}a the exact transmission $|t(\omega)|^2$
of a cavity coupled to two Gaussian ensembles and verified that the position of the
peaks are fitted by the eigenenergies computed in the discrete case. Moreover, we have
superimposed the transmission resulting from the exact calculation and from the discrete
model, as it can be seen in fig.\ref{dark}b after focusing on the central peak of the
transmission pattern: the excellent agreement between the two plots fully validates the
effective approach. This central peak corresponds to the eigenstate $\ket{\psi_d}$
resulting from the coupling between the cavity mode and the antisymmetric state
$\ket{{\cal A}} = (\ket{G_1,S_2}-\ket{S_1,G_2})/\sqrt{2}$, its expression being
$\ket{\psi_d}=(i \delta \ket{1,G_1,G_2} + \Omega \sqrt{2} \ket{0,{\cal
A}})/\sqrt{\delta^2+2\Omega^2}$. When $ \delta \gg \Omega$, the excitation is mostly in
the cavity, and mostly in the matter field in the opposite case. This change of nature
clearly appears in the narrowing of the peak while lowering $\delta$, as it can be seen
in the figure, and confirmed by the expression for its linewidth $\Gamma_d =( \delta^2
\kappa + 2\Omega^2 \gamma ) /(\delta^2+2\Omega^2)$. Note that this modelling might
explain some recent experimental results \cite{Ong}, in which a superconducting cavity
is strongly coupled to a inhomogeneous ensemble of NV centers of spin 1. Because of the
geometrical strain, the transitions $\ket{m_S=0}\rightarrow \ket{m_S=1}$ and
$\ket{m_S=0}\rightarrow \ket{m_S=-1}$ are splitted, which can be modeled by two
ensembles of emitters of different central frequencies. The visible presence of a narrow
peak at the cavity frequency explains qualitatively the effect discussed above.

Coming back to the general case of two distinct ensembles, the state $\ket{\psi_d}$
could provide a new type of quantum memory as mentioned in the beginning of this
Section. The protocol consists in feeding the cavity mode with a single photon while the
ensembles are largely detuned, thus preparing the state $\ket{1,G_1,G_2}$, then
adiabatically transferring the excitation to $\ket{\psi_d}$ by slowly lowering $\delta$.
Yet the ensembles cannot be brought to resonance with the mode as it would be the case
for two discrete emitters. As it appears in fig.\ref{dark}c, the effective model breaks
down when $\delta \sim \Delta$. At this point indeed, the distributions of emitters
start to spectrally overlap with the central peak, leading to its broadening. This
yields a minimal linewidth $\Gamma_d \sim \gamma + (\Delta^2/2\Omega^2) \kappa$,
allowing to typically reduce the cavity losses by $(\Omega/\Delta)^2$. Here again the
ratio $(\Omega/\Delta)^2$ appears as a major figure of merit for devices based on
inhomogeneous ensembles strongly coupled to cavities.

\section{Conclusion}

We have shown that if an inhomogeneous distribution of emitters is strongly coupled to a
cavity, the ensemble can be treated as a single effective emitter collectively coupled
to the mode, whose relaxation is governed by single emitter's properties, provided that
their spectral distribution decreases faster than $1/\omega^2$. This effect called "cavity protection" offers promising
perspectives in the framework of quantum information with solid state integrable
devices, in particular regarding the implementation of long lived high fidelity quantum
memories. These results are quite general, and can fruitfully be applied to numerous
important physical systems, ranging from semiconductor emitters coupled to optical
cavities, to ensembles of spins in circuit QED.

\begin{acknowledgments}
The authors gratefully thank Z. Kurucz, K. M{\o}lmer, G. Nogues, J. Claudon, J.P.
Poizat, M.F. Santos and D. Est\`eve for all the fruitful exchanges. This work was
supported by NanoSci-ERA consortium and by the EU under ERANET project LECSIN, by the
Nanosciences Foundation of Grenoble, and the ANR project CAFE. ID acknowledges the
CAPES. ID, SP and AA thank the Center for Quantum Technologies of Singapore.

\end{acknowledgments}

\appendix

\section{Dynamics} \label{appendix:Dynamics}

In this part, we establish the link between the complex transmission of the cavity, and
the evolution of the system if the mode $a$ is initially fed with a single photon. This evolution is governed by
the set of equations (\ref{input}) written in the free frame ($\omega=0$).
The input fields are in the vacuum, the state of the system is
$\ket{1,G}=a^\dagger(0) \ket{0}$ where $\ket{0}$ is the ground state of the total
system. We are interested in the quantities $\langle a(t) a^\dagger (0) \rangle$ and
$\langle b_k(t) a^\dagger(0) \rangle$, which represent the probability amplitude of the
excitation in the cavity mode and in each emitter respectively, as it will appear later.
The average values are taken in state $\ket{0}$. We get

\begin{equation}\begin{split}
&\langle \dot{a}(t) a^\dagger(0) \rangle = - \roundbraket{\kappa/2 + i\omega_0 }  \langle a(t) a^\dagger(0) \rangle+  \sum_k g_k \langle b_k(t) a^\dagger(0) \rangle \\
&\langle \dot{b_k}(t) a^\dagger(0)\rangle = - \roundbraket{\gamma/2 + i\omega_k }
\langle b_k(t) a^\dagger(0) \rangle - g_k \langle a(t) a^\dagger(0) \rangle \,
\end{split}\end{equation}

Defining the vector $\ket{\psi}$ of coordinates $( \langle a(t) a^\dagger (0) \rangle,
..., \langle b_k(t) a^\dagger(0) \rangle, ...)$, its evolution simply follows the
Schr\"odinger like equation  $\hbar {\displaystyle \frac{d}{dt}} \ket{\psi}(t) =  -i
H_{eff} \ket{\psi}(t)$, with

\begin{equation}
  H_{eff}/\hbar =  \left(
           \begin{array}{cccc}
             \tilde\omega_0 & i g_1 &  i g_2 & \ldots \\
             -i g_1               & \tilde\omega_1 &     & \\
             -i g_2               &                 & \tilde\omega_2 &\\
             \vdots            &                 &                &  \ddots   \\
           \end{array}
         \right)\, .
\end{equation}

We have used the complex frequencies for the cavity $\tilde{\omega}_0$ and for the
emitters  $\tilde{\omega}_k$ defined above. Note that these results are in full
agreement with the ones obtained in the Green function formalism by Kurucz et al
\cite{Zoltan}. It appears that the dynamics of the problem can be modeled with the
effective Hamiltonian $H_{eff}$. In particular, one can define an effective evolution
operator $O(t) = e^{i H_{eff} t /\hbar} O e^{-i H_{eff} t /\hbar}$, such that $\langle
a(t) a^\dagger (0) \rangle = \bra{0} a(0) e^{-i H_{eff} t /\hbar} a^\dagger(0) \ket{0}$.
This quantity can be rewritten $\bra{1,G} e^{-i H_{eff} t /\hbar } \ket{1,G}$,
justifying that we talk of the probability amplitude of the excitation in the cavity
mode, starting from the initial state $\ket{1,G}$. The problem is solved using e.g.
standard Laplace transform method. Defining $ \mathcal{L} \roundbraket{f(t)} = F(s) =
\int_{0}^\infty \exp(s t) f(t) dt $, we have

\begin{equation}\begin{split}\label{Laplace sys}
\ket{\psi(t)} = \mathcal{L}^{-1}  \roundbraket{ (s+i H_{eff} /\hbar)^{-1}  \ket{\psi(0)}
} \, ,
\end{split}\end{equation}

where we have used the Laplace transform property: $ \mathcal{L} \{ {\displaystyle
\frac{d}{dt}} \ket{\psi(t)} \} = s \ket{\Psi(s)} - \ket{\psi(0)} $. We finally define
$t_1(s) = \bra{1,G} (s + i H_{eff} /\hbar)^{-1}  \ket{1,G}$. Inverse Laplace transform
of this coefficient gives back the quantity $\bra{1,G} e^{-i H_{eff} /\hbar t}
\ket{1,G}$. We easily get

\begin{equation}\begin{split}
t_1(s) =  \frac{1}{s + i \tilde\omega_0  + \sum_{k} \frac{g_k^2}{s + i \tilde\omega_k} }
\, , \label{eq:t1(s) = sum_k}
\end{split}\end{equation}

From eq.(\ref{eq:t(omega)in-out}) and eq.(\ref{eq:t1(s) = sum_k}), we finally write the
link between the transmission coefficient in amplitude $t(\omega)$ and the coefficient
$t_1(s)$ characterizing the dynamics of the system,

\begin{equation}\begin{split}
t(\omega) = - \frac{\kappa}{2} t_1(-i \omega)  \, .
\end{split}\end{equation}

This establishes the relation between the amplitude $\alpha_1(t) = \bra{1,G} e^{-i H_{eff} t
/\hbar } \ket{1,G}$ and the transmission $t(\omega)$ as:

\begin{equation}\begin{split}\label{eq:linktransm a1}
& \int_{0}^\infty \alpha_1(t) e^{i \omega t} dt  =  - \frac{2}{\kappa} t(\omega) \, .\\
\end{split}\end{equation}

One can use the method exposed above to compute the expression of the probability
amplitude for a state initially prepared in $ \ket{\psi^0_+(\delta)}$, namely $\bra{\psi^0_+(\delta)} e^{-i
H_{eff} t /\hbar}  \ket{\psi^0_+(\delta)}$ studied in Section \ref{sec:QM}. In general, we can decompose it as

\begin{equation}\begin{split}
& \bra{\psi^0_+(\delta)} U_{eff}(t) \ket{\psi^0_+(\delta)} = \\ %
& = \cos^2(\theta/2) \bra{1,G} U_{eff}\ket{1,G} + \sin^2(\theta/2) \bra{0,S} U_{eff} \ket{0,S} +  \\ %
&+ i \sin(\theta/2) \cos(\theta/2) \roundbraket{ \bra{0,S} U_{eff} \ket{1,G} - \bra{1,G} U_{eff} \ket{0,S}} \\ %
& = \cos^2(\theta/2) \alpha_1(t) + \sin^2(\theta/2) \alpha_2(t) + \\
& + i \sin(\theta/2) \cos(\theta/2) \roundbraket{\alpha_3(t) - \alpha_4(t) } \, ,
\end{split}\end{equation}
where $U_{eff}(t) \equiv  e^{-i H_{eff} t /\hbar}$ .

\bigskip

We need only to calculate the four matrix elements $\alpha_i(t)$. Defining $t_i(s) =
\mathcal{L} (\alpha_i(t))$ we obtain in the case of a continuous distribution,
\begin{eqnarray} \label{eq:t2}\begin{split}
t_2(s)  &=  - \frac{ W( i s )}{ \Omega^2 } t_1(s) (s + i \tilde{\omega_o} ) \\ 
t_3(s) &=t_1( s) \frac{i W( i s)}{\Omega}  \\
t_4(s) &= - t_3(s)  \, .
\end{split}\end{eqnarray}

\section{$W(\omega)$ for specific distributions} \label{appendix:specificW}

We now evaluate the function $W(\omega)$ for all the specific continua analyzed in the
paper. This function allows the evaluation of the complex transmission using $t(\omega)
= (\kappa / 2i) (\omega-\omega_0 + i\kappa/2-W(\omega) )^{-1}$, but also appears in
other formulae.

\subsection*{Gaussian}

The Gaussian distribution writes $\rho(\omega) = \frac{\sqrt{ \ln 2} }{ \Delta \sqrt{\pi}} e^{- ( \omega^2 \ln2) /
\Delta^2}$. $W(\omega)$ is thus,

\begin{equation}\begin{split}
W_G(\omega) = \frac{1}{i} \frac{\sqrt {\ln2} \; \Omega^2}{ \Delta} \sqrt \pi
 \roundbraket{\frac{i}{\pi} \int_{-\infty}^{\infty} \frac{ d\omega' e^{-\omega'^2} }{
\roundbraket{ \frac{\omega + i \gamma /2}{ \Delta / \sqrt {\ln2}} - \omega' } } } \, .
\end{split}\end{equation}

Remembering that
\begin{equation} \label{eq:W(z)} \begin{split}
\frac{i}{\pi} \int_{-\infty}^{\infty} d\omega' \frac{e^{-\omega'^2} }{ z - \omega' } =
e^{-z^2}  \operatorname{erfc}(-i z)
\end{split}\end{equation}
where $\operatorname{erfc}$ is the complex complementary error function,
it comes
\begin{equation}\begin{split}
W_G(\omega) = -i \frac{\sqrt {\ln2} \; \Omega^2}{ \Delta} \sqrt \pi e^{-(\frac{\omega +
i \gamma /2}{ \Delta / \sqrt {\ln2} })^2} \operatorname{erfc} \roundbraket
{-i\frac{\omega + i \gamma /2}{ \Delta / \sqrt {\ln2} } } \, .
\end{split}\end{equation}

\subsection*{Rectangular}

In the case of a rectangular distribution, the density of emitter is $\rho(\omega) =
\frac{1}{ \Delta } (\Theta(\omega - \Delta /2) - \Theta(\omega + \Delta /2) )$ we have

\begin{equation}\begin{split}
W_R(\omega) = \frac{2 \Omega^2}{i \Delta} \operatorname{ArcTan}\roundbraket{
\frac{\Delta }{\gamma - 2 i \omega} } \, .
\end{split}\end{equation}

\subsection*{Lorentzian}

The density of emitter is $\rho(\omega) = \frac{\Delta /2}{\pi} \frac{1}{(\Delta /2)^2 +
\omega^2} $ , thus

\begin{equation}\begin{split}
W_L(\omega) = \frac{\Omega^2}{\omega + i\gamma/2 + i\Delta /2}\, . %
\end{split}\end{equation}

From the equation above we see that for Lorentz distribution we do not achieve cavity
protection, i.e. the inhomogeneous broadening always contributes as if it were
homogeneous.

\section{Development with finite $\gamma$} \label{appendix:DevelopmentFinitegamma}

We start rewrite $W(\omega)$ from eq.(\ref{eq:Wdefinition}) as
\begin{equation}\begin{split}
W(\omega) = \Omega^2  \int_{-\infty}^{\infty} d\omega' \frac{\omega'^2}{\omega'^2 + \gamma^2} \frac{\rho(\omega'+\omega)}{\omega'} - \\ %
- i \pi \Omega^2 \int_{-\infty}^{\infty} d\omega' \frac{\gamma}{\pi(\omega'^2 + \gamma^2) }  \rho(\omega + \omega')  \, . %
\end{split}\end{equation}

The integrands contain products of a function of width $\gamma$ and another with width
$\Delta$. If $\gamma \ll \Delta$, the integrals take the form:

\begin{equation}\begin{split} \label{eq:Wlowgamma}
W(\omega) &= \Omega^2 P\!\!\! \int_{-\infty}^\infty  \frac{\rho(\omega') d \omega' }{ \omega - \omega' } -  \\
&-i \Omega^2 \roundbraket{\pi \rho(\omega) + \frac{\gamma}{2} \; P\!\!\! \int_{-\infty}^\infty  \frac{ \rho(\omega') d \omega' }{ (\omega - \omega')^2 } }   \, . %
\end{split}\end{equation}

We are interested in the development of $W(\omega)$ near the poles of the transmission function
in the absence of inhomogeneous broadening, namely $\omega \sim \Omega$.
Denoting $r=\omega'/\omega$, and using the identity $\sum r^k = 1/ (1-r)$, we find :

\begin{equation}
\begin{split}
W(\omega) &= \frac{\Omega^2}{\omega} \roundbraket{1 + \sum_{k=1}^\infty \frac{\mu_k}{\omega^k} - i \pi \rho(\omega)}  - \\%
& - i \frac{\Omega^2}{\omega^2} \frac{\gamma}{2} \roundbraket{ 1 + \sum_{k=1}^\infty (k+1) \frac{\mu_k}{\omega^k}  }  \, ,%
\end{split}\end{equation}
where $\mu_k$ is the $k$-th moment of the distribution $\rho(\omega)$ about its origin
\begin{equation} \begin{split}
  \mu_k \equiv \int_{- \infty}^\infty d\omega \rho(\omega) \omega^k  \, .%
\end{split}\end{equation}

Note that this development is only valid if $\omega \gg \omega'$, which is the case in the present study
as $\omega \sim \Omega \gg \Delta > \omega'$.
From the normalization and considering only symmetric distributions, we have $\mu_0=1$ and $\mu_1=0$. $\mu_2$ gives the first
non-zero correction and it is typically proportional to the square of the FWHM (as an example, $\mu_2 = \Delta^2 / ( 2 \ln 2)$ in the case of a Gaussian distribution).
To first non-zero correction we have:

\begin{equation}
\begin{split}
W(\omega) &= \frac{\Omega^2 }{ \omega} (1 + \mu_2/ \omega^2 ) - i \roundbraket{\frac{\gamma}{2}\frac{\Omega^2}{\omega^2}+ \pi \Omega^2 \rho(\omega) } \\%
&= \frac{\Omega^2 (1 + \mu_2/ \omega^2 ) }{\omega + i\gamma/2} - i\pi \Omega^2 \rho(\omega) \label{eq:W=i(ga + rho) + Re} \, , %
\end{split}\end{equation}
where we have used $\Omega \gg \gamma$.
One easily infers the modifications to the transmission poles induced
by inhomogeneous broadening. They are located at

\begin{equation} \label{eq:appendixPoles}
\omega_\pm = \pm \Omega \sqrt{1 + \mu_2/ \Omega^2 - \left(\frac{\kappa+2\pi \rho(\Omega)\Omega^2-\gamma}{4\Omega}\right)^2}.
\end{equation}

Their width check $\Gamma = \frac{\kappa+\gamma+2\pi\Omega^2 \rho(\Omega)}{2}$,
in correspondence with what stated in Sec. \ref{sec:TranmissionProperties}.
Note that this procedure is only valid for distribution with well defined moments. This
is not the case of the Lorentzian, nevertheless $W(\omega)$ can be exactly evaluated in
this case. The exact calculations for the 3 cases taken under consideration are the
subject of appendix \ref{appendix:specificW}.

\section{Two ways to obtain the temporal evolution} \label{appendix:DynamicsVsFano}

We have found two ways to evaluate  $\alpha_1(t) = \bra{1,G} e^{-i H_{eff} t /\hbar }
\ket{1,G}$, the first at Appendix \ref{appendix:Dynamics} uses a Laplace-Fourier
transform of $-t(\omega) / (\kappa/2)$ the second uses the standard Fourier transform of
$ 2 \pi \Omega^2 \; \rho(\omega) \left| t(\omega) / (\kappa/2 ) \right|^2$ for $\kappa,
\gamma \ra 0$ as in Sec. \ref{sec:exactEingenV}. The first way is more general in the
sense that it can include emitter and cavity radiative losses, the second describe a
reversible process originated in a Hamiltonian evolution. We now show that both ways
coincide when we disregard losses.

From Appendix \ref{appendix:Dynamics} we have
\begin{equation}\begin{split}\label{eq:E1}
& \int_{0}^\infty \alpha_1(t) e^{i \omega t} dt  =  t_1(-i \omega) \, ,\\
\end{split}\end{equation}
where, if $\gamma,\kappa \ra 0$
\begin{equation}\begin{split} \label{eq:E2}
& t_1(-i \omega) = \frac{i}{\omega - \omega_0 - \Omega^2 P\!\! \int  \frac{\rho(\omega') d \omega' }{\omega - \omega' } %
+ i \pi \Omega^2 \rho(\omega) } \, .
\end{split}\end{equation}

We now take the real part of eqs.(\ref{eq:E1} , \ref{eq:E2}) , yielding
\begin{equation}\begin{split} \label{eq:E3}
&\Re \curlybraket{ \int_{0}^\infty \alpha_1(t) e^{i \omega t}  dt } = \\
&= \Re \curlybraket{ \frac{i}{\omega - \omega_0 - \Omega^2 P\!\! \int \frac{\rho(\omega') d \omega' }{\omega - \omega' } + i \pi \Omega^2 \rho(\omega) } }\\ %
&= \pi  \Omega^2 \rho(\omega) |t_1|^2 \, ,
\end{split}\end{equation}

if we consider time reversibility of the lossless dynamics we have $\alpha_1(-t) =
(\alpha_1(t))^*$ and thus
\begin{equation}\begin{split} \label{eq:E4}
& 2 \Re \curlybraket{ \int_{0}^\infty \alpha_1(t) e^{i \omega t}  dt } =  \, \\
& = \int_{0}^\infty (\alpha_1(t) e^{i \omega t} + \alpha_1^*(t) e^{-i \omega t})  dt  \\ %
& = \int_{-\infty}^\infty \alpha_1(t) e^{i \omega t} dt  \, .\\
\end{split}\end{equation}

Eq.(\ref{eq:E3}) and eq.(\ref{eq:E4}) together give
\begin{equation}\begin{split}
\int_{-\infty}^\infty \alpha_1(t) e^{i \omega t} dt  = 2 \pi \Omega^2 \; \rho(\omega) \left| t_1 \right|^2  \, , %
\end{split}\end{equation}
which is precisely what we find applying the inverse Fourier transform in
eq.(\ref{eq:1G=F(a^2)}). Note we had to use the time-reversibility which is only valid
in the lossless case.

\end{document}